\documentclass[prd,showpacs,amsmath,amssymb]{revtex4}
\usepackage{graphicx,color}
\begin{document}



\title{
Effects of hadronic loops on the direct CP violation of $B_{c}$}

\author{Xiang Liu$^1$ and Xue-Qian Li$^2$}


\vspace*{1.0cm}

\affiliation{$^{1}$Department of physics, Peking University,
Beijing, 100871, China}

\affiliation{$^{2}$Department of physics, Nankai University,
Tianjin, 300071, China}

\vspace*{1.0cm}

\date{\today}
\begin{abstract}
{ It is well known that the final state interaction plays an
important role in the decays of $B$-meson. The contribution of the
final state interaction which is supposed to be long-distance
effects, to the concerned processes can interfere with that of the
short-distance effects produced via the tree and/or loop diagrams
at quark-gluon level. The interference may provide a source for
the direct CP violation $\mathcal{A}_{CP}$ in the process
$B_{c}^{+}\to D^{0}\pi^{+}$. We find that a typical value of
$\mathcal{A}_{CP}$ when the final state interaction effect is
taken into account can be about $-22\%$ which is different from
that without the final state interaction effect. Therefore, when
we extract information on CP violation from the data which will be
available at LHCb and the new experiments in $B$-factories, the
contribution from the final state interaction must be included.
This study may be crucial for searching new physics in the future.
}
\end{abstract}

\pacs{11.30.Er, 13.75.Lb, 13.25.Ft} \maketitle


\section{introduction}

One of the most intriguing goals in the high energy physics is to
look for new physics beyond the Standard Model (SM) via heavy
hadron production and decay processes. The reason is that new
physics which generally has a higher energy scale may be observed
at the processes involving heavy flavors. Among all the possible
quantities which are experimentally measurable, CP violation
provides a more sensitive window to the new physics effects.
Direct CP violation at $B$-physics has been observed by the Babar
and Belle collaborations \cite{Babar-cp-B,Belle-cp-B}, which is
indeed a great success after confirmation of non-zero
$\epsilon'/\epsilon$ at K-systems. Another promising place to
study CP violation is the meson $B_c$, which is composed of
different heavy flavors.

Since the CDF Collaboration observed $B_{c}$ meson in the
semileptonic decay $B_{c}\to J/\psi+l+\nu$ \cite{CDF}, studies on
$B_{c}$ have drawn great interests from both theorists and
experimentalists of high energy physics. Decays of $B_{c}$ can be
realized via $b-$decay, $\bar c-$decay and annihilation of $b$ and
$\bar c$ \cite{LM}. Many theoretical works have been dedicated to
study the decays of $B_{c}$
\cite{korner,formfactor-1,formfactor-2,decay}. A relatively
complete discussion about its spectrum, production and decays was
presented in a review \cite{review}. Because of the specific
characteristics of its decay modes, the direct CP violation is an
important observable which may provide valuable information
towards the mechanism governing the transition and probably
unveils a trace to the new physics beyond the SM.

In this work, we are just looking for a new source for the direct
CP violation in $Bc$ decays. The direct CP violation is caused in
general, by an interference among at least two channels which have
the same final state, but different weak and strong phases. The CP
quantity  $A_{CP}$ is proportional to
$$A_{CP}={2|A_1||A_2|\sin(\theta_1-\theta_2)\sin(\alpha_1-\alpha_2)\over
|A_1|^2+|A_2|^2+2|A_1|A|_2|\cos(\theta_1-\theta_2)\cos(\alpha_1-\alpha_2)},$$
where $A_1,A_2$ are the amplitudes of the two distinct channels
and $\theta_1,\theta_2$, $\alpha_1,\alpha_2$ are their strong
phases and weak phases respectively.

These phase differences coming from either quark level or hadron
level. At the quark level the strong phase difference usually
occurs via the absorptive part of the loops involved in the
calculation. The strong phase may also occur at the hadron level.
As a matter of fact, it is well known in the kaon system. When one
studies the direct CP violation, i.e. $\epsilon'/\epsilon$, the
phase shifts in the $\pi\pi$ scattering provide the strong phase
which are necessary to result in CP violation. But recently most
of the works to study direct CP violation concentrate on the
strong phase induced by the absorptive part of the loops.
Especially, the strong phase is coming from the absorptive part of
the penguin diagram(s) which contribute along with the tree
diagram to the amplitude. In that case the CP violation is induced
by the interference between the contribution of the tree diagram
and that of penguin.

The total width is related to $|A_1+A_2|^2$. If one of the
amplitudes is much smaller than the other one, the width should be
only depend on the larger one, say $A_1$, thus one can ignore the
smaller one when he is calculating the decay width. However, even
though $|A_2|\ll |A_1|$, the numerator of $A_{CP}$ is proportional
to their product, so one cannot ignore the smaller contribution,
otherwise he would get null CP asymmetry.

In that case, obviously the contribution from the penguin diagram
is much smaller than that from the tree diagram, so that if only
the decay width is needed, one can completely ignore the
contribution of the penguin. However, for evaluating the CP
violation, he by no means can dismiss the penguin contribution.


In the SM, the weak phase originates from the
Cabibbo-Kabayashi-Maskawa matrix, and the strong phase is induced
by the absorptive part of loops. At the quark-gluon level which is
responsible for the short distance effects, the strong phases may
originate from the absorptive part of loops, for example, the
penguin diagrams. On the other aspect, the final state interaction
(FSI) plays an important role in $B$-physics, as fully discussed
in literature \cite{HY-Chen}.  At the short distance, the direct
CP violation usually is caused by an interference between the
tree-level contribution and the loop-induced one because they have
different weak and strong phases (in fact the tree diagrams do not
contribute a strong phase). Therefore an interference of the
long-distance contribution with the short-distance ones may change
the theoretical prediction on the CP violation. In fact, the FSI
effect is extensively applied to the discussion of the CP
violation of $B$ and $D$ decays \cite{Du-Li,B-cp,FSI}.

Indeed, by the quantum field theory, the lagrangian can be a
combination of various pieces and each of them corresponds to
different processes. For our transition matrix element
$M=_{out}\langle f|i\rangle_{in}$, one has
$$M= \langle f|L^{(1)}_{PQCD}+T[L_{had}L^{(2)}_{PQCD}]|i\rangle,$$
where $L_{PQCD}^{(1),(2)}$ corresponds to the lagrangian which
includes QCD and weak  or electromagnetic interactions, the
superscripts (1) and (2) denote the lagrangians which can lead to
different final states, whereas $L_{had}$ is the lagrangian at
hadron level. Then we further write the matrix element as
\begin{equation}\label{amplitude}
M=\langle f|L_{PQCD}^{(1)}|i\rangle+\sum_n\langle
f|L_{had}|n\rangle\langle n|L_{PQCD}^{(2)}|i\rangle,
\end{equation}
where the intermediate states $|n\rangle$ are a complete set of
hadrons with proper quantum numbers and the matrix element
$\langle f|L_{had}|n\rangle$ is just the hadronic scattering
process and corresponds to the hadronic loops in our work.

Generally, the long-distance effects due to the FSI refer to the
re-scattering of the intermediate hadrons which emerge at the
direct decays, into the concerned final state and it is depicted
by the term $\langle n|L_{PQCD}^{(2)}|i\rangle$. In these channels
with the intermediate hadronic intermediate states may have
different weak phases from that of the short-distance production
channel occurring at quark-gluon-level. In the re-scattering
processes, phase shifts exist due to strong interaction and thus
can offer strong phases. And an extra strong phase which is
definitely different from that induced by the quark-level loops,
occurs from the hadron re-scattering processes $\langle
f|L_{had}|n\rangle$.

Since the loop contribution is suppressed by the loop integration,
generally the second term of the above equation is smaller than
the first one which we may call it as the "tree" level
contribution( but maybe not the tree diagram in the common sense).

The traditional PQCD calculation only takes care of the first term
$\langle f|L_{PQCD}^{(1)}|i\rangle$ and  $\langle
n|L_{PQCD}^{(2)}|i\rangle$ in the second one, but leaves the part
$\langle f|L_{had}|n\rangle$ to be dealt with in other theories,
for example, the chiral lagrangian and etc. at the hadron level.
This picture is clearly depicted in Cheng's paper \cite{HY-Chen}.
It indicates that unless the "tree" contribution (i.e. the first
term) is suppressed by some mechanism, the first term corresponds
to the direct process, so that is always dominating for the total
amplitude. If we only need to consider the total decay width, the
second term may contribute a smaller portion (sometimes it might
be enhanced by some mechanism, but generally is much smaller).
However, as we deal with the CP violation and need at least two
different channels, their interference enforces us not to abandon
this term even though it might be much smaller than the first one.
Indeed, one may argue that the loop diagram, such as penguin can
also contribute a strong phase and interfere with the tree
contribution to result in a direct CP violation, the loop
contribution may have a similar order as we considered here and
possibly even smaller. At least as we state above, we are looking
for a possible source of CP violation in $Bc$ decays, i.e. the
hadronic loops may contribute strong phases and cause sizable
effects on CP violation as our numerical results given in the
paper indicate.

Therefore we would say that the PQCD framework works well, but we
instead are looking for a supposed-to-be smaller effect which can
result in observable CP violation. If there is a small
double-counting (could be), that is because the wavefunction
adopted in the calculation is not well defined. In fact because
$|n\rangle$ generally are not the same as $|f\rangle$, the
double-counting does not appear.

At present the direct CP violation in $B_{c}$ decays due to short
distance contribution has been studied by many authors
\cite{CP-1,kt-chao,CP-2,CP-3,cai-dian lu}. But so far, the studies
of the FSI effects on the direct CP Violation of $B_{c}$ are
absent. In this work, by taking into account the long distance
contribution caused by the FSI, we would re-evaluate the direct CP
violation in the decays of $B_c$. Namely, we add a new
contribution to the amplitude which has different strong and weak
phases from that of short distance contributions which were
calculated by many authors, thus their interference will
significantly change the value of CP asymmetry in decays of $B_c$.


Indeed, before one can claim a discovery of new physics, he must
exhaust all possibilities which the SM can provide. Therefore this
work is also serving for the purpose and to determine if the FSI
can result in a sizable contribution to the direct CP violation of
$B_c$.

We choose the channel $B_{c}^{+}\to D^{0}\pi^{+}$ which should be
one of the dominant decay modes of $B_c$. In this channel, there
exist hadronic intermediate states which are mainly composed of
$D^{(*)+}$ and $J/\psi$.


This paper is organized as follow. We present the formulation
about $B_{c}^{+}\to D^{(*)+}J/\psi\to D^{0}\pi^{+}$ in \ref{sec2}.
Then we present our numerical results. The last section is a short
conclusion and discussion.

\section{formulation}\label{sec2}
The effective Hamiltonian related to $B_{c}$ decays is \cite{EH}
\begin{eqnarray}
\mathcal{H}_{eff}&=&\frac{G_{F}}{\sqrt{2}}V_{cb}V_{cd}^{\dag}\Big\{\mathcal{C}_{1}^{b}(\mu)
(\bar{c}b)_{V-A})(\bar{d}c)_{V-A}
+\mathcal{C}_{2}^{b}(\mu)(\bar{d}b)_{V-A})(\bar{c}c)_{V-A}\Big\}\nonumber\\&&
+\frac{G_{F}}{\sqrt{2}}V_{ub}V_{ud}^{\dag}\Big\{\mathcal{C}_{1}^{b}(\mu)
(\bar{u}b)_{V-A})(\bar{d}u)_{V-A}+\mathcal{C}_{2}^{b}(\mu)(\bar{d}b)_{V-A})(\bar{u}u)_{V-A}\Big\},
\end{eqnarray}
where the subscript $V-A$ denotes the left-chiral current
$\gamma^{\mu}(1-\gamma^{5})$. $\mathcal{C}_{1,2}^{b}(\mu)$ denote
the Wilson coefficients .

Firstly we calculate the transition amplitude of $B_{c}\to
D^{(*)+}J/\psi$ at the quark level and the hadronization would be
described by a few phenomenological parameters.  The definitions
of the relevant hadronic matrix elements are
\begin{eqnarray}
\langle0|\mathcal{J}_{\mu}|\mathcal{P}(k)\rangle&=&-if_{\mathcal{P}}k_{\mu},\\
\langle0|\mathcal{J}_{\mu}|\mathcal{V}(k,\epsilon)\rangle&=&f_{\mathcal{V}}\epsilon_{\mu}m_{\mathcal{V}},
\end{eqnarray}
where $f_{\mathcal{P}}$ and $f_{\mathcal{V}}$ respectively stand
for leptonic decay constants of pseudoscalar and vector mesons.
$k_{\mu}$ is the four-momentum of the concerned hadron and
$\epsilon_{\mu}$ denotes the polarization of the vector meson. One
has
$\mathcal{J}_{\mu}=\bar{q}_{1}\gamma_{\mu}(1-\gamma_{5})q_{2}$.

In addition, the hadronic matrix elements of $B_{c}$ transiting into
two mesons  can be expressed in terms of a few form factors as
\cite{EH}
\begin{eqnarray} \langle
\mathcal{P}(k_{2})|\mathcal{J}_{\mu}|B_{c}(k_1)\rangle&=&P_{\mu}f_{+}(Q^{2})+{Q}_{\mu}f_{-}(Q^{2}),\\
{1\over
i}\langle\mathcal{V}(k_{2},\epsilon)|\mathcal{J}_{\mu}|B_{c}(k_1)\rangle&=&\frac{\epsilon_{\nu}^{*}}
{m_{1}+m_{2}}\big\{i\varepsilon^{\mu\nu\alpha\beta}
P_{\alpha}Q_{\beta}F_{V}(Q^2)-g^{\mu\nu}(P\cdot
Q)F_{0}^{A}(Q^2)\nonumber\\&&+P^{\mu}P^{\nu}F_{+}^{A}(Q^{2})+Q^{\mu}P^{\nu}F_{-}^{A}(Q^{2})\big\}
\end{eqnarray}
with $P_{\mu}=(k_{1}+k_{2})_{\mu}$ and
$Q_{\mu}=(k_{1}-k_{2})_{\mu}$. With the above formulas, we obtain
\begin{widetext}
\begin{eqnarray}
&&\mathcal{M}[B^{+}_{c}(p)\to
D^{+}(p_1)J/\psi(p_2)]\nonumber\\&=&
\frac{iG_F}{\sqrt{2}}V_{cb}V_{cd}^{\dag} \Big\{
a_{1}f_{D}\frac{p_{1\sigma}}{m_{B_{c}}+m_{\psi}}\Big[
-g^{\sigma\lambda}(p+p_{2})\cdot(p-p_{2})F^{A}_{0}(q_{1}^{2})+(p+p_{2})^{\sigma}(p+p_{2})^{\lambda}F_{+}^{A}(q_{1}^2)
\nonumber\\&&+(p-p_{2})^{\sigma}(p+p_{2})^{\lambda}F_{-}^{A}(q_{1}^{2})
\Big] +a_{2}f_{\psi}m_{\psi}\big[(p+p_{1})^{\lambda}f_{+}(q_{2}^{2})
+(p-p_{1})^{\lambda}f_{-}(q_{2}^2) \big] \Big\},
\end{eqnarray}
and
\begin{eqnarray}
&&\mathcal{M}[B^{+}_{c}(p)\to D^{*+}(p_1)J/\psi(p_2)]\nonumber\\&&=
\frac{iG_F}{\sqrt{2}}V_{cb}V_{cd}^{\dag} \Big\{
a_{1}f_{D^*}m_{D^{*}}\frac{i}{m_{B_{c}}+m_{\psi}}\Big[
i\varepsilon^{\sigma\omega\tau\delta}(p+p_{2})_{\tau}(p-p_{2})_{\delta}F_{V}(q_{1}^2)\nonumber\\&&-
g^{\sigma\omega}(p+p_{2})\cdot(p-p_{2})F^{A}_{0}(q_{1}^{2})+(p+p_{2})^{\sigma}(p+p_{2})^{\omega}F_{+}^{A}(q_{1}^2)
\nonumber\\&&+(p-p_{2})^{\sigma}(p+p_{2})^{\omega}F_{-}^{A}(q_{1}^{2})
\Big] + a_{2}f_{\psi}m_{\psi}\frac{i}{m_{B_{c}}+m_{D^{*}}}\Big[
i\varepsilon^{\omega\sigma\tau\delta}(p+p_{1})_{\tau}(p-p_{1})_{\delta}F_{V}(q_{2}^2)\nonumber\\&&-
g^{\omega\sigma}(p+p_{1})\cdot(p-p_{1})F^{A}_{0}(q_{2}^{2})+(p+p_{1})^{\omega}(p+p_{1})^{\sigma}F_{+}^{A}(q_{2}^2)
+(p-p_{1})^{\omega}(p+p_{1})^{\sigma}F_{-}^{A}(q_{2}^{2})
\Big]\nonumber\\
\end{eqnarray}
\end{widetext}
with $q_{1}=p-p_{2}$ and $q_{2}=p-p_{1}$. The values of $a_{1,2}$
will be given in next subsection.

\subsection{Absorptive part of hadronic loop for $B_{c}^{+}\to D^{(*)+}J/\psi\to D^{0}\pi^{+}$}

\begin{figure}[htb]
\begin{center}
\begin{tabular}{cccccccc}
\scalebox{0.85}{\includegraphics{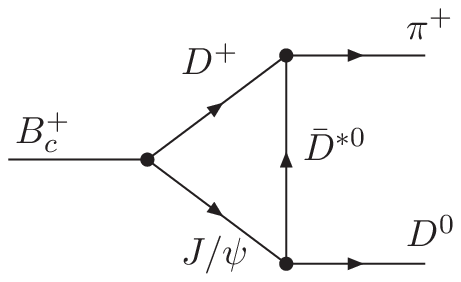}}&\scalebox{0.85}{\includegraphics{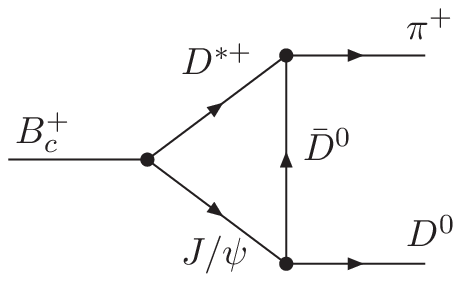}}&
\scalebox{0.85}{\includegraphics{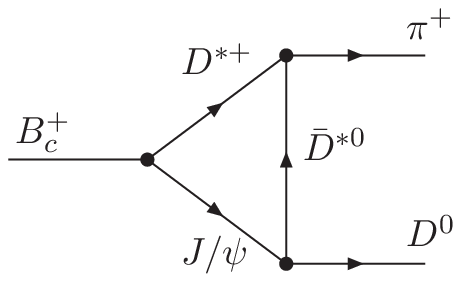}}
\\(a)&(b)&(c)
\end{tabular}
\end{center}
\caption{The final state interaction contributions to
$B^{+}_{c}\to D^{0}\pi^{+}$.\label{hard}}
\end{figure}

Now let us turn to evaluate the contribution from the
long-distance effects which occur at the hadron level. The
diagrams shown in  Fig. \ref{hard} depict sequent processes
$B_{c}^{+}\to D^{(*)+}J/\psi\to D^{0}\pi^{+}$.

The effective Lagrangian at the hadronic level is suggested to be
in the following forms as \cite{lagrangian}
\begin{eqnarray}
\mathcal{L}_{\mathcal{D}^{*}\mathcal{D}\pi}&=&ig_{\mathcal{D}^{*}\mathcal{D}\pi}
(\mathcal{D}^{*}_{\mu}\partial^{\mu}{\pi}\bar{\mathcal{D}}-\mathcal{D}\partial^{\mu}{\pi}\bar{\mathcal{D}}^{*}_{\mu}),\\
\mathcal{L}_{\mathcal{D}^{*}\mathcal{D}^{*}\pi}&=&-g_{\mathcal{D}^{*}\mathcal{D}^{*}\pi}\varepsilon^{\mu\nu\alpha\beta}
\partial_{\mu}\mathcal{D}^{*}_{\nu}{\pi}\partial_{\alpha}\bar{\mathcal{D}}_{\beta}^{*},\\
\mathcal{L}_{\psi \mathcal{D}\mathcal{D}}&=&ig_{\psi \mathcal{D}
\mathcal{D}}\psi_{\mu}(\partial^{\mu}\mathcal{D}\bar{\mathcal{D}}-\mathcal{D}\partial^{\mu}\bar{\mathcal{D}}),\\
\mathcal{L}_{\psi \mathcal{D}^{*}\mathcal{D}}&=&-g_{\psi
\mathcal{D}^{*}
\mathcal{D}}\varepsilon^{\mu\nu\alpha\beta}\partial_{\mu}\psi_{\nu}(\partial_{\alpha}D_{\beta}^{*}
\bar{\mathcal{D}}+\mathcal{D}\partial_{\alpha}\bar{\mathcal{D}}^{*}_{\beta})\nonumber\\
\end{eqnarray}
with $\pi={\mbox{\boldmath $\tau$}}\cdot {\mbox{\boldmath $\pi$}}$,
where fields $\mathcal{D}^{(*)}$ and $\bar{\mathcal{D}}^{(*)}$ are
defined as $\mathcal{D}^{(*)}=(D^{(*)0},D^{(*)+})$ and
$\bar{\mathcal{D}}^{(*)T}=(\bar{D}^{(*)0},\bar{D}^{(*)-})$.

\begin{widetext}
The process shown in Fig. 1 (a) is $B_{c}^{+}\to
D^{+}(p_{1})J/\psi(p_2)\to \pi^{+}(p_{3})D^{0}(p_{4})$ where
$D^{*0}$ is exchanged at t-channel, and its amplitude reads
\begin{eqnarray}\label{a1}
Abs^{(a)}&=&\frac{1}{2}\int\frac{d^{3}p_{1}}{(2\pi)^{3}2E_{1}}\frac{d^{3}p_{2}}{(2\pi)^{3}2E_{2}}(2\pi)^{4}\delta^{4}
(m_{B_{c}}-p_{1}-p_{2})\mathcal{M}[B^{+}_{c}(p)\to
D^{+}(p_{1})J/\psi(p_{2})]\nonumber\\&&\times\big[-g_{D^{*}D\pi}ip_{3}^{\xi}\big]
\big[-ig_{J/\psi D^{*}D}\varepsilon^{\mu\nu\alpha\beta}(-ip_{2\mu})
iq_{\alpha}\big](-g_{\lambda\nu} + \frac{p_{2\lambda}
p_{2\nu}}{m_{\psi}^{2}} )\nonumber \\
&&\times (-g_{\beta\xi}+\frac{q_{\beta} q_{\xi}}{m_{D^{*}}^{2}} )
\frac{i}{q^2-m_{D^{*}}^2}\mathcal{F}^2[q^2,m_{D^{*}}^2].
\end{eqnarray}
Obviously, the conservation of angular momentum demands the
contribution from Fig. 1 (a) to be zero.

The amplitude corresponding to the process of $B_{c}^{+}\to
D^{*+}(p_{1})J/\psi(p_2)\to \pi^{+}(p_{3})D^{0}(p_{4})$ where
$D^{0}$ is exchanged at t-channel reads as
\begin{eqnarray}\label{a2}
Abs^{(b)}&=&\frac{1}{2}\int\frac{d^{3}p_{1}}{(2\pi)^{3}2E_{1}}\frac{d^{3}p_{2}}{(2\pi)^{3}2E_{2}}(2\pi)^{4}\delta^{4}
(m_{B_{c}}-p_{1}-p_{2})\mathcal{M}[B^{+}_{c}(p)\to
D^{*+}(p_1)J/\psi(p_2)]\nonumber\\&&\times\big[-g_{D^{*}D\pi}(-ip_{3}^{\xi})
 \big] \big[-g_{\psi D D}
(ip_{4}-iq)^{\mu}\big] (-g_{\sigma\xi}+\frac{p_{1\sigma}
p_{1\xi}}{m_{D^{*}}^{2}} )\nonumber\\&&\times(-g_{\omega\mu}
+\frac{p_{2\omega} p_{2\mu}}{m_{\psi}^{2}}
)\frac{i}{q^2-m_{D}^2}\mathcal{F}^2[q^2,m_{D}^2].
\end{eqnarray}

For Fig. 1 (c), $B_{c}^{+}\to D^{*+}(p_{1})J/\psi(p_2)\to
\pi^{+}(p_{3})D^{0}(p_{4})$ where $D^{*0}$ is exchanged at
t-channel, the amplitude is
\begin{eqnarray}\label{a3}
Abs^{(c)}&=&\frac{1}{2}\int\frac{d^{3}p_{1}}{(2\pi)^{3}2E_{1}}\frac{d^{3}p_{2}}{(2\pi)^{3}2E_{2}}(2\pi)^{4}\delta^{4}
(m_{B_{c}}-p_{1}-p_{2})\mathcal{M}[B^{+}_{c}(p)\to
D^{*+}(p_1)J/\psi(p_2)]\nonumber\\&&\times
[-ig_{D^{*}D^{*}\pi}\varepsilon^{\mu\nu\alpha\beta}(-ip_{1\mu})
(-iq_{\alpha})] \big[-ig_{J/\psi
D^{*}D}\varepsilon^{\xi\lambda\kappa\rho}(-ip_{2\xi})
iq_{\kappa}\big]\nonumber\\&& \times(-g_{\beta\rho}
+\frac{q_{\beta} q_{\rho}}{m_{D^{*}}^{2}} ) (-g_{\sigma\nu}
+\frac{p_{1\sigma} p_{1\nu}}{m_{D^{*}}^{2}} )(-g_{\omega\lambda}
+\frac{p_{2\omega} p_{2\lambda}}{m_{\psi}^{2}}
)\frac{i}{q^2-m_{D^{*}}^2}\mathcal{F}^2[q^2,m_{D^{*}}^2].
\end{eqnarray}
\end{widetext}

In the above amplitudes, $q=p_{3}-p_{1}$, and
$\mathcal{F}(q^2,m_{i})$ etc. denote the form factors which
compensate the off-shell effects of mesons at the effective
vertices and may be described by the possible pole structures
\cite{HY-Chen}
\begin{eqnarray}
\mathcal{F}(q^2,m_{i})=\bigg(\frac{\Lambda^{2}-m_{i}^2
}{\Lambda^{2}-q^{2}}\bigg)^{n},
\end{eqnarray}
where $\Lambda$ is a phenomenological parameter to be determined.
As $q^2\to 0$ the form factor becomes a number. If $\Lambda\gg
m_{i}$, it becomes a unity. As $q^2\rightarrow\infty$, the form
factor approaches to zero. It reflects the fact that as the
distance between the mesons becomes very small, their inner
structures would overlap and the whole picture of hadron
interaction breaks down. Hence the form factor vanishes at large
$q^2$ and effectively plays a role to cut off the ultraviolet
divergence. The expression of $\Lambda$ is suggested to be
\cite{HY-Chen}
\begin{eqnarray}
\Lambda(m_{i})=m_{i}+\alpha \Lambda_{QCD},\label{parameter}
\end{eqnarray}
where $m_{i}$ denotes the mass of the exchanged meson and $\alpha$
is a phenomenological parameter. In this work, we adopt the dipole
form factor $\mathcal{F}(q^2,m_{i})={(\Lambda^{2}-m_{i}^2)^2
}/{(\Lambda^{2}-q^{2})^2}$.

\subsection{The dispersive part of $B_{c}^{+}\to D^{(*)+}J/\psi\to
D^{0}\pi^{+}$}

In the above subsection, the absorptive part of the triangle
diagram to the amplitude of the sequent process $B_{c}^{+}\to
D^{(*)+}J/\psi\to D^{0}\pi^{+}$ can be easily obtained from the
integrals (\ref{a1}), (\ref{a2}) and (\ref{a3}). The dispersive
part of $B_{c}^{+}\to D^{(*)+}J/\psi\to D^{0}\pi^{+}$ can be
related to the absorptive part via the dispersive relation
\cite{shifman,HY-Chen}
\begin{eqnarray}
{Dis}[B_{c}^{+}\to
D^{0}\pi^{+}]&&=\frac{1}{\pi}\int^{\infty}_{s_{1}}\frac{Abs[B_{c}^{+}\to
D^{0}\pi^{+}]}{s-m_{B_c}^{2}}ds.\label{dispersive relation}
\end{eqnarray}
However, the cutoff which is phenomenologically introduced and the
complicated integral in eq. (\ref{dispersive relation}) would
cause unavoidable uncertainties to the dispersive part. In some of
the former works, for estimating the decay width, the contribution
of dispersive part was assumed to be small comparing with that of
the absorptive part and ignored. However, for the direct CP
violation, we must estimate the dispersive part and determine the
strong phase induced by the triangle diagram, otherwise the strong
phase would be exactly $\pi/2$.

In this work, adopting the method in our previous work \cite{liu},
we obtain the dispersive part of $B_{c}^{+}\to D^{(*)+}J/\psi\to
D^{0}\pi^{+}$ by directly calculating the triangle where the
intermediate hadrons are not on their mass shells. The amplitudes
corresponding to the process of $B_{c}^{+}\to
D^{*+}(p_{1})J/\psi(p_2)\to \pi^{+}(p_{3})D^{0}(p_{4})$ where
$D^{0}$ or $D^{*0}$ are exchanged are
\begin{eqnarray}
{Dis}^{(b)}&=&\int\frac{d^4
q}{(2\pi)^4}\mathcal{M}[B^{+}_{c}(p)\to
D^{*+}(p_1)J/\psi(p_2)]\big[-g_{D^{*}D\pi}(-ip_{3}^{\xi})
 \big] \big[-g_{\psi D D}
(ip_{4}-iq)^{\mu}\big] \nonumber\\&&\times(-g_{\sigma\xi}
)(-g_{\omega\mu}
)\frac{i}{p_{1}^2-m_{D}}\frac{i}{p_{2}-m_{J/\psi}}
\frac{i}{q^2-m_{D}^2}\mathcal{F}^2[q^2,m_{D}^2],\label{real1}
\end{eqnarray}
and
\begin{eqnarray}
Dis^{(c)}&=&\int\frac{d^4 q}{(2\pi)^4}\mathcal{M}[B^{+}_{c}(p)\to
D^{*+}(p_1)J/\psi(p_2)][-ig_{D^{*}D^{*}\pi}
\varepsilon^{\mu\nu\alpha\beta}(-ip_{1\mu})
(-iq_{\alpha})]\nonumber\\&&\big[-ig_{J/\psi
D^{*}D}\varepsilon^{\xi\lambda\kappa\rho}(-ip_{2\xi})
iq_{\kappa}\big](-g_{\beta\rho}
) (-g_{\sigma\nu}
)(-g_{\omega\lambda}
)\frac{i}{p_{1}^2-m_{D}}\frac{i}{p_{2}-m_{J/\psi}}\nonumber\\&&\times
\frac{i}{q^2-m_{D^{*}}^2}\mathcal{F}^2[q^2,m_{D^{*}}^2].\label{real2}
\end{eqnarray}
Due to the existence of the  dipole form factors
$\mathcal{F}^2[q^2,m_{D}^2]$ and $\mathcal{F}^2[q^2,m_{D^{*}}^2]$
the ultraviolet behavior of the triangle loop integration is
benign. These form factors play an equivalent role to the
$\Lambda-$related terms introduced in the Pauli-Villas
renormalization scheme \cite{Izukson,peskin}. Because the final
expressions of eqs. (\ref{real1}) and (\ref{real2}) are
complicated, we would collect some useful formulas in appendix.

\subsection{Direct CP violation}


The observable direct CP violation is defined as
\begin{eqnarray}
\mathcal{A}_{CP}&=&\frac{|\mathcal{M}|^2-|\overline{\mathcal{M}}|^2}
{|\mathcal{M}|^2+|\overline{\mathcal{M}}|^2}
\end{eqnarray}
with
\begin{eqnarray*}
&&\mathcal{M}=\mathcal{M}^{dir}(B_{c}^+\to
D^{0}\pi^{+})+\mathcal{M}^{FSI}(B_{c}^+\to D^{0}\pi^{+}),\\
&&\overline{\mathcal{M}}=\mathcal{M}^{dir}(B_{c}^-\to
\bar{D}^{0}\pi^{-})+\mathcal{M}^{FSI}(B_{c}^-\to
\bar{D}^{0}\pi^{-}),
\end{eqnarray*}
where $\mathcal{M}^{dir}(B_{c}^{+}\to D^{0} \pi^{+})$ and
$\mathcal{M}^{dir}(B_{c}^{-}\to \bar{D}^{0} \pi^{-})$ were
calculated in the approach of PQCD by the many authors
\cite{cai-dian lu} and they are written as
\begin{eqnarray}
&&\mathcal{M}^{dir}(B_{c}^{+}\to D^{0} \pi^{+})=
V_{u}(T_{u}+P)[1-ze^{i(-\gamma+\delta)}],\\
&&\mathcal{M}^{dir}(B_{c}^{-}\to \bar{D}^{0} \pi^{-})=
V_{u}^{*}(T_{u}+P)[1-ze^{i(\gamma+\delta)}]
\end{eqnarray}
with
$$z=\bigg|\frac{V_{c}}{V_{u}}\bigg|\bigg|\frac{T_{c}+P}{T_{u}+P}\bigg|\;\;
{\rm and} \;\;\;\delta=\arg\Big[\frac{T_{c}+P}{T_{u}+P}\Big],$$
where $V_{u}=V_{ud}V_{ub}^*$ are the Cabibbo-Kabayashi-Maskawa
entries,
$|V_{c}/V_{u}|=\frac{\lambda}{1-\lambda^2/2}|V_{cb}/V_{ub}|$. The
values of $T_{u,c}$, $P$, $\gamma$, $\lambda$, $z$ and $\delta$
are given in Ref. \cite{cai-dian lu}, which are listed in Table
\ref{PQCD}.

The amplitude of $B_{c}^+\to D^{0}\pi^+$ induced by the FSI effect
which is denoted by the subscript "$FSI$" is:
\begin{eqnarray}
\mathcal{M}^{FSI}(B_{c}^+\to
D^{0}\pi^{+})=Dis+i\sum_{j=a,b,c}Abs^{(j)},
\end{eqnarray} and
\begin{eqnarray}
{\mathcal{M}}^{FSI}(B_{c}^+\to
D^{0}\pi^{+})=\mathcal{M}^{FSI}(B_{c}^-\to \bar{D}^{0}\pi^{-}).
\end{eqnarray}

\begin{center}
\begin{table}[htb]
\begin{tabular}{c|c||c|cccc}\hline
$T_{u}$&$22.621+0.863i$&$\delta$&$123^{\circ}$ \\
$T_{c}$&$-0.83+3.57i$&$\gamma$&$55^{\circ}$\\
$P$&$-0.474-1.722i$&$|\frac{V_{ub}}{V_{cb}}|$&0.085\\
$z$&$0.28$&&\\\hline
\end{tabular}\caption{These values are taken from  Ref. \cite{cai-dian lu}.
Here $T_{u,c}$ and $P$ are in unit of $10^{-3}$ GeV. In this work,
we need multiply a factor
${\sqrt{m_{B_c}^5}G_F}/{\sqrt{2|\mathbf{k}|}}\sim 4.83\times
10^{-4}$ to $T_{u,c}$ and $P$, because the formula for the decay
widths adopted in this work takes a different normalization from
that in Ref. \cite{cai-dian lu}.\label{PQCD}}
\end{table}
\end{center}

\section{numerical results}

The input parameter set which we are going to use in this work,
includes: $m_{B_{c}}=6.286$ GeV, $m_{J/\psi}=3.097$ GeV,
$m_{D^{+}}=1.869$ GeV, $m_{D^{*+}}=2.01$ GeV, $m_{D^{0}}=1.865$
GeV \cite{PDG}; $f_{\psi}=405\pm17$ MeV \cite{PDG};
$f_{D}=222.6\pm 16.7^{+2.8}_{-3.4}$ MeV,
$f_{D^*}=245\pm20^{+3}_{-2}$ MeV \cite{fd}. $V_{ud}=0.974$,
$V_{cd}=0.230$, $V_{cb}=0.0416$ \cite{PDG}. $g_{D^{*}D\pi}=17.3$,
$g_{D^{*}D^{*}\pi}=8.9$ GeV$^{-1}$, $g_{DD\psi}=7.9$, $g_{D^*
D\psi}=4.2$ GeV$^{-1}$ \cite{coupling}; $a_{1}=1.14$,
$a_{2}=-0.20$ \cite{korner}.
$V_{ub}=A\lambda^{3}(\rho-i\eta)=0.00218-0.00335i$. The
wolfenstein parameters of CKM matrix elements: $\lambda=0.2272$,
$A=0.818$, $\bar{\rho}=0.221$ and $\bar{\eta}=0.340$ with
$\bar{\rho}=\rho(1-\frac{\rho}{2})$ and
$\bar{\eta}=\eta(1-\frac{\rho}{2})$. $G_F=1.16637\times 10^{-5}$
GeV$^{-2}$
\cite{PDG}.

The form factors in processes $B_{c}\to D^{(*)}$ and $B_{c}\to
J/\psi$ possess pole structures \cite{formfactor-1,formfactor-2}
\begin{eqnarray}
F(q^2)=\frac{F(0)}{1-a\zeta+b\zeta^2}
\end{eqnarray}
with $\zeta=q^2/m_{B_{c}}^{2}$. The values of $F(0)$, $a$ and $b$
are evaluated by some authors and for readers' convenience, we
list their results in Table \ref{table-1}.
\begin{table}[htb]
\begin{center}
\begin{tabular}{c||c|cc|ccccc} \hline
&&$f_{+}$&$f_{-}$&$F_{+}^{A}$&$F_{-}^{A}$&$F_{0}^{A}$&$F_{V}$\\\hline\hline
&$F(0)$&0.189&-0.194&-&-&-&-\\\cline{2-8}
&$a$&2.47&2.43&-&-&-&-\\\cline{2-8}
\raisebox{2.5ex}[0pt]{$D$}&$b$&1.62&1.54&-&-&-&-\\\hline\hline
&$F(0)$&-&-&0.158&-0.328&0.284&0.296\\\cline{2-8}
&$a$&-&-&2.15&2.40&1.30&2.40\\\cline{2-8}
\raisebox{2.5ex}[0pt]{$D^*$}&$b$&-&-&1.15&1.51&0.15&1.49\\\hline\hline
&$F(0)$&-&-&0.66&-1.13&0.68&0.96\\\cline{2-8}
&$a$&-&-&1.13&1.23&0.59&1.24\\\cline{2-8}
\raisebox{2.5ex}[0pt]{$J/\psi$}&$b$&-&-&-0.067&0.006&-0.483&-0.002\\\cline{1-8}
\end{tabular}
\end{center}\caption{The values of $F(0)$, $a$ and $b$ in the form factors of
$B_{c}\to D^{(*)}$ and $B_{c}\to J/\psi$
\cite{formfactor-1,formfactor-2}. \label{table-1}}
\end{table}

\begin{figure}[htb]
\begin{center}
\begin{tabular}{c}
\scalebox{0.85}{\includegraphics{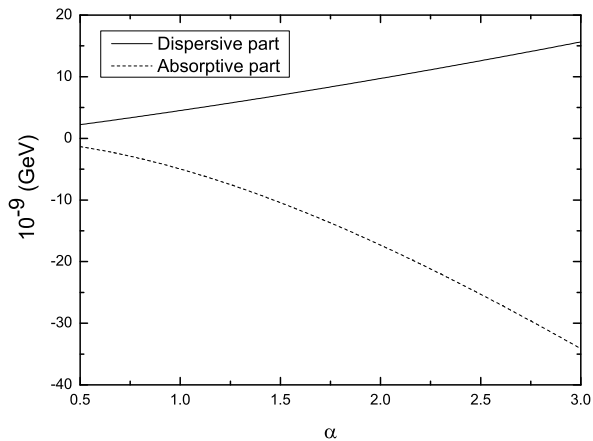}}
\end{tabular}
\end{center}\caption{The dispersive part and absorptive part of the amplitudes
of $B_{c}^{+}\to D^{(*)+}J/\psi\to D^{0}\pi^{+}$.\label{real-img}}
\end{figure}

\begin{center}
\begin{figure}[htb]
\begin{tabular}{c}
\scalebox{0.85}{\includegraphics{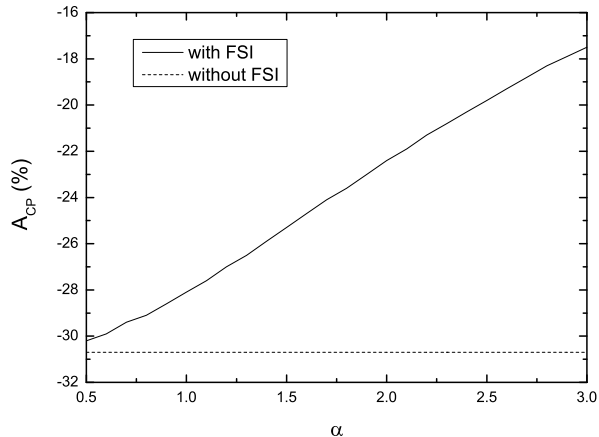}}
\end{tabular}
\caption{The direct CP violation. Solid line and Dash line
correspond to the direct CP with FSI effect and without FSI effect
respectively. \label{CP}}
\end{figure}
\end{center}

\begin{center}
\begin{table}[htb]

\begin{tabular}{c||cccccc}\hline
$\alpha$&0.5&1.0&1.5&2.0&2.5&3.0\\\hline
 $\mathcal{A}_{CP}$&$-30.2\%$&$-28.1\%$&$-25.3\%$&$-22.4\%$&$-19.8\%$&$-17.5\%$\\\hline
\end{tabular}\caption{The typical values of $\mathcal{A}_{CP}$}
\end{table}
\end{center}

In Fig. \ref{real-img}, we plot the dispersive part and absorptive
part of the amplitudes of $B_{c}^{+}\to D^{(*)+}J/\psi\to
D^{0}\pi^{+}$ versus $\alpha$ which is allowed to vary within
$\alpha=0.5\sim 3$. In Fig. \ref{CP}, we also list
$\mathcal{A}_{CP}$ with several typical values of $\alpha$. For a
clear comparison, in this figure, we also give the value of
$\mathcal{A}_{CP}$ calculated in Ref. \cite{cai-dian lu} by the
PQCD approach which is purely induced by the short distance
contribution (without considering the FSI). For clarity we also
list some typical values of $\mathcal{A}_{CP}$ with various
$\alpha$ in Table III.

\section{Discussion and Conclusion}

Recently the direct CP violation in $B$-decays has been observed
and it is expected to open a window for exploring new physics
beyond the SM by which all theorists and experimentalists feel
very inspired. Obviously, investigation of direct CP violation at
$B_c$ decays would be of special interests because it is composed
of two heavy flavors and may be more sensitive to new physics. On
other aspect, before one can claim to find a trace of new physics,
he must exhaust all possibilities in the framework of the SM. As
indicated in literature, the FSI play an important role in B
decays, therefore one has a full reason to expect that it is also
significant at $B_C$ decays. In this work, we carefully study the
contribution of the FSI to the direct CP violation via its
interference with the contribution from the short-distance effects
which are induced by the tree and loop diagrams. Concretely, in
this work, we calculate the amplitudes  for $B_{c}^{+}\to
D^{0}\pi^{+}$ via sequent processes $B_{c}^{+}\to
D^{(*)+}J/\psi\to D^{0}\pi^{+}$ and determine its strong and weak
phases.

Here we need to add some interpretation about application of PQCD.
Even though we indicate the significance of the FSI for evaluating
direct CP violation in Bc decays, their absolute contribution is
much smaller than that from the direct process which is calculated
in the framework of PQCD. Therefore if only the decay width of Bc
is needed, one can ignore the contribution from the hadronic
re-scattering, but as the CP violation is concerned as we see
above, its contribution might be significant.

Our numerical results indicate that the typical value of
$\mathcal{A}_{CP}$ with FSI effect is about $-22\%$, which is
different from the value $-30.7\%$ estimated in the PQCD approach
without FSI  \cite{cai-dian lu}. On other aspect, one can also
observe from Fig.3 that the effect of the hadronic re-scattering
on $A_{CP}$ may change quite diversely depending on the input
parameter. Especially, as one adopts $\alpha=0.5$, $A_{CP}$ is
about $-30.2\%$ which only slightly deviates from the value
obtained in the framework of PQCD, however, as $\alpha=3$, (even
though $\alpha=3$ seems too large to be very reasonable, this
effective coupling indeed can exceed 1 for hadron interaction, in
fact, in some applications its value is set to be very large for
fitting data) $A_{CP}$ would change to $-17.5\%$ obviously
deviates from the value of PQCD. It indicates that the
contribution of FSI to $A_{CP}$ is of opposite sign with that from
the quark loops and the cancellation may cause remarkable effects
when one analyzes the data achieved in a rather precise
measurement. Thus our conclusion is that  the contribution from
the FSI is not negligible.

In the future experiments, especially the LHCb, a great amount of
data on $B_c$ will be accumulated and one may have a possibility
to measure the direct CP violation of $B_c$. If non-zero
$\mathcal{A}_{CP}$ (almost definitely yes) is well measured, one
can look for a trace of new physics by comparing the measured
value with the theoretical result. When one compares the data
which will be available at LHCb and/or other experiments with
theoretical predictions, the contribution from the FSI must be
included. This observation may be crucial for searching new
physics in the future.

 \vfill

\section*{Acknowledgments}

X.L. thanks Prof. Shi-Lin Zhu and B. Zhang for useful discussion.
This project was supported by the National Natural Science
Foundation of China under Grants 10421503, 10625521, 10475042 and
10705001, Key Grant Project of Chinese Ministry of Education (No.
305001), the China Postdoctoral Science foundation (No.
20060400376) and the Ph.D. Programs Foundation of Ministry of
Education of China (No. 20020055016).  \vfill

\section*{Appendix}
Some useful formulas in the calculation of eqs. (\ref{real1}) and
(\ref{real2}):
\begin{eqnarray*}
&&\int\frac{d^4
q}{(2\pi)^4}\frac{1}{(p_{1}^2-m_{1}^2)(p_2-m_2)(q^2-m^2)}\Big(\frac{\Lambda^2-m^2}{q^2-\Lambda^2}\Big)^4\\&=&
\frac{i}{(4\pi)^2}\int^{1}_{0}dx\int^{1-x}_{0}dy\Big\{
\frac{(\Lambda^2-m^2)y}{\Delta^2(m_1,m_2,\Lambda)}
+\frac{1}{\Delta(m_1,m_2,\Lambda)}-\frac{1}{\Delta(m_1,m_2,m)}\nonumber\\&&
-\frac{(-\Lambda^4-m^4+2m^2\Lambda^2)y^2}{\Delta^3(m_1,m_2,\Lambda)}
 +\frac{(\Lambda^6-3m^2\Lambda^4+3m^4
\Lambda^2-m^6)y^3}{\Delta^4(m_1,m_2,\Lambda)}\Big\}.
\end{eqnarray*}

\begin{eqnarray*}
&&\int\frac{d^4
q}{(2\pi)^4}\frac{l^2}{(p_{1}^2-m_{1}^2)(p_2-m_2)(q^2-m^2)}\Big(\frac{\Lambda^2-m^2}{q^2-\Lambda^2}\Big)^4\\&=&
\frac{i}{(4\pi)^2}\int^{1}_{0}dx\int^{1-x}_{0}dy\Big\{
\frac{-2y(\Lambda^2-m^2)}{\Delta(m_1,m_2,\Lambda)}
+2\ln\Big[\frac{\Delta(m_1,m_2,\Lambda)}{\Delta(m_1,m_2,m)}\Big]
\nonumber
\\&&+\frac{(-\Lambda^4-m^4+2m^2\Lambda^2)y^2}{\Delta^2(m_1,m_2,\Lambda)}
 -\frac{2y^3(\Lambda^6-3m^2\Lambda^4+3m^4
\Lambda^2-m^6)}{3\Delta^3(m_1,m_2,\Lambda)}\Big\}.
\end{eqnarray*}

\begin{eqnarray*}
&&\int\frac{d^4
q}{(2\pi)^4}\frac{l^4}{(p_{1}^2-m_{1}^2)(p_2-m_2)(q^2-m^2)}\Big(\frac{\Lambda^2-m^2}{q^2-\Lambda^2}\Big)^4\\&=&
\frac{i}{(4\pi)^2}\int^{1}_{0}dx\int^{1-x}_{0}dy\Big\{
6y(\Lambda^2-m^2)\ln\Big[\frac{1}{\Delta(m_1,m_2,\Lambda)}\Big]\nonumber\\&&
-4y(\Lambda^2-m^2)+6\Big[\Delta(m_1,m_2,\Lambda)\ln\Big(\frac{1}{\Delta(m_1,m_2,\Lambda)}\Big)\nonumber
\\&&-\Delta(m_1,m_2,m)\ln\Big[\frac{1}{\Delta(m_1,m_2,m)}\Big]
-\frac{(3y^2\Lambda^4-m^4+2m^2\Lambda^2)}{\Delta(m_1,m_2,\Lambda)}
\nonumber\\&&+\frac{y^3(\Lambda^6-3m^2\Lambda^4+3m^4
\Lambda^2-m^6)}{\Delta^2(m_1,m_2,\Lambda)}\Big\}.
\end{eqnarray*}
Here
$$q=l-p_4 x+p_3 (1-x-y),$$ and
\begin{eqnarray*}
\Delta(a,b,c)&=&a^2(1-x-y)+b^2x+c^2y
+m^2_3(x^2+y^2-x-y+2xy)+m^2_4(x^2-x)\nonumber\\&&+p_3\cdot p_4
(2x^2+2xy-2x).
\end{eqnarray*}


\end{document}